\documentclass[aps,citeautoscript,prb,twocolumn]{revtex4}
\usepackage{amsfonts}
\usepackage{amsmath}
\usepackage{amssymb}
\usepackage{graphicx}

\def\kbar{\protect\@kbar}
\def\@kbar{\relax \bgroup
\def\@tempa{\hbox{\raise.73\ht0
\hbox to0pt{\kern.25\wd0\vrule width.5\wd0 height.1pt
depth.1pt\hss}\box0}}\mathchoice{\setbox0\hbox{$\displaystyle
k$}\@tempa}{\setbox0\hbox{$\textstyle
k$}\@tempa}{\setbox0\hbox{$\scriptstyle
k$}\@tempa}{\setbox0\hbox{$\scriptscriptstyle k$}\@tempa}\egroup}

\begin{document}

\title{\textbf{Classical and Quantum Transport in One-Dimensional
Periodically Kicked Systems}}
\author{Itzhack Dana}
\affiliation{Department of Physics, Bar-Ilan University, Ramat-Gan
52900, Israel}

\begin{abstract}
This paper is a brief review of classical and quantum transport phenomena,
as well as related spectral properties, exhibited by one-dimensional
periodically kicked systems. Two representative and fundamentally different
classes of systems will be considered, those satisfying the classical
Kolmogorov-Arnol'd-Moser scenario and those which not. The experimental
realization of some of these systems using atom-optics methods will be
mentioned.
\newline


\end{abstract}

\maketitle

\newpage

\begin{center}
\textbf{I. INTRODUCTION}
\end{center}

During the last four decades, there has been much interest in the problem of
\textquotedblleft Quantum Chaos\textquotedblright , i.e., understanding the
properties and dynamics of quantum systems whose classical Hamiltonian
counterparts are nonintegrable and exhibit chaos. In developing this
understanding, there was a natural attempt to focus, at least at a first
stage, on model systems that are as simple as possible but still feature
some typical behavior of more complex and/or \textquotedblleft
realistic\textquotedblright\ nonintegrable systems. The minimal number of
degrees of freedom required by a nonlinear Hamiltonian system to be
nonintegrable is \textquotedblleft 1.5\textquotedblright , namely one
dimension (1D) with time dependence. A simple time dependence is a periodic
one and a very simple periodic dependence is that of a periodic delta
function. This corresponds to 1D periodically \textquotedblleft
kicked\textquotedblright\ systems. The most attractive feature of these
systems, and also of kicked systems in higher dimensions, is that both their
classical Poincar\'{e} map and quantum map in one time period can be written
explicitly in closed form. This fact already makes it possible to derive
several exact results and/or to make rigorous statements about the
properties of classical and quantum kicked systems.

The many theoretical investigations of these relatively simple systems, some
of which eventually became paradigmatic models in the field of Quantum
Chaos, have led to the discovery of an unexpected rich variety of
fascinating classical and quantum transport phenomena. These systems have
also proved to be quite realistic. In fact, during the last two decades,
several of the quantum phenomena have been experimentally realized using
atom-optics methods with cold atoms or Bose-Einstein condensates. In most
cases, the experimental results agreed well with the theoretical
predictions. Some phenomena were first discovered experimentally and later
explained theoretically.

This paper is a brief review of classical and quantum transport phenomena,
as well as related spectral properties, exhibited by representative classes
of 1D periodically kicked systems. This review will include significant
contributions made by Paul Brumer and co-workers to the field, see Secs. III
and IVD. The paper is organized as follows. In Secs. II and III, we consider
kicked systems satisfying the classical Kolmogorov-Arnol'd-Moser (KAM)
scenario. Generalized versions of paradigmatic models in this class of
systems, the kicked rotor and the kicked particle, will be treated in some
detail in Sec. II; variants of these models, the modulated kicked rotors,
will be the subject of Sec. III. In Sec. IV, we consider the fundamentally
different class of \textquotedblleft non-KAM\textquotedblright\ kicked
systems, represented by generalized kicked charged particles in a uniform
magnetic field or, equivalently, generalized kicked harmonic oscillators; a
small sub-class of these systems is exactly equivalent to generalized
versions of the paradigmatic \textquotedblleft kicked Harper
models\textquotedblright . Conclusions are presented in Sec. V.

\begin{center}
\textbf{II. KAM SYSTEMS: KICKED ROTOR AND KICKED PARTICLE}\\[0pt]
{\ }\\[0pt]
\textbf{A. Classical Kicked Rotor, KAM Scenario, Chaotic Normal and
Anomalous Diffusion}
\end{center}

The classical periodically kicked rotor (KR), or Chirikov-Taylor system \cite{bc}, is defined, in its generalized version and in dimensionless scaled
variables, by the Hamiltonian:
\begin{equation}
H=\frac{L^{2}}{2}+kV(\theta )\sum_{t=-\infty }^{\infty }\delta (t^{\prime
}-t),  \label{gkr}
\end{equation}
where $L$ is angular momentum, $\theta $ is angle, $k$ is a nonintegrability
parameter, $V(\theta )$ is a general $2\pi $-periodic potential, $t^{\prime
} $ is the usual (continuous) time, and $t$ is a \textquotedblleft
discrete\textquotedblright\ time taking all integer values. The classical
map for (\ref{gkr}), from time $t^{\prime }=t-0$ to time $t^{\prime }=t+1-0$
, is:
\begin{equation}
L_{t+1}=L_{t}+kf(\theta _{s}),\ \ \ \theta _{t+1}=\theta _{t}+L_{t+1}\ \text{
mod}(2\pi ),  \label{gsm}
\end{equation}
where the subscripts indicate the times above at which $(L,\theta )$ are
evaluated and $f(\theta )=-dV/d\theta $ is the force function. For $V(\theta
)=\cos (\theta )$ or $f(\theta )=\sin (\theta )$, Eqs. (\ref{gsm}) give the
famous \textquotedblleft \emph{standard map}\textquotedblright\ \cite{bc,jg}
, a prototypical Hamiltonian map exhibiting the celebrated
Kolmogorov-Arnol'd-Moser (KAM) scenario \cite{bc,mmp}: As $k$ is gradually
increased from $0$, local chaotic layers develop around \textquotedblleft
stochastic resonances\textquotedblright\ which extend \textquotedblleft
horizontally\textquotedblright\ from $\theta =0$ to $\theta =2\pi $ but are
bounded in the \textquotedblleft vertical\textquotedblright\ ($L$) direction
by \textquotedblleft horizontal\textquotedblright\ KAM tori. For $k>k_{
\mathrm{c}}\approx 0.9716$ \cite{jg}, all the KAM tori break into
\textquotedblleft cantori\textquotedblright\ \cite{icp} with gaps, so that
the local chaotic layers merge into a single global chaotic region extending
from $L=-\infty $ to $L=\infty $. Then, unbounded \emph{chaotic diffusion} 
\cite{bc,rrw,df,sd1}\ takes place in the $L$ direction, characterized by the
diffusion coefficient
\begin{equation}
D=\lim_{t\rightarrow \infty }\frac{\left\langle
(L_{t}-L_{0})^{2}\right\rangle }{2t},  \label{D}
\end{equation}
where $\left\langle \cdot \right\rangle $ denotes average over an ensemble $
\left\{ (L_{0},\theta _{0})\right\} $ of initial conditions in the chaotic
region. For $k$ sufficiently larger than $k_{\mathrm{c}}$, the standard map
features generalized periodic orbits, the so-called \emph{\textquotedblleft
accelerator modes\textquotedblright } \cite{sd1,sd2,id1,id2}, satisfying
\begin{equation}
L_{m}=L_{0}+2\pi w,\ \ \theta _{m}=\theta _{0}\ \text{mod}(2\pi ),\ \ 
\label{am}
\end{equation}
where $m$ is the minimal period and $w\neq 0$ is some integer. \ Clearly,
the orbit (\ref{am}) performs ballistic motion in the $L$ direction with
mean acceleration $2\pi w/m$. If this orbit is stable, it generates a chain
of $m$ islands embedded in the chaotic sea and all the points within these
islands perform ballistic motion with the same mean acceleration:
\begin{equation}
\left\langle (L_{mt}-L_{0})^{2}\right\rangle \approx 4\pi ^{2}w^{2}t^{2},
\label{bm}
\end{equation}
where the average is over an initial ensemble within the islands. A chaotic
orbit will usually stick to the boundaries of the island chain for a long
time, performing ballistic motion during this time, it will then return to
the chaotic sea where it will diffuse for some time, and it will eventually
stick again to the island-chain boundaries. The net result of these
transitions over a very long time is an \emph{anomalous} chaotic diffusion 
\cite{sd1,sd2,id1,id2}:
\begin{equation}
\left\langle (L_{mt}-L_{0})^{2}\right\rangle \propto t^{\mu },\ \ \ 1<\mu <2,
\label{sd}
\end{equation}
where the average is again over an initial ensemble in the chaotic region
and $\mu $ is the anomalous-diffusion exponent. The latter ranges between
the values of $1$ [normal diffusion (\ref{D})] and $2$ [ballistic motion 
(\ref{bm})]. It corresponds therefore to a \emph{\textquotedblleft
superdiffusion\textquotedblright }, which appears to be the most
well-established kind of Hamiltonian anomalous diffusion (there exists, however, a
rare case \cite{sk} of Hamiltonian subdiffusion, $\mu <1$). It must
be remarked that the existence of accelerator modes (\ref{am}) and the
associated chaotic superdiffusion (\ref{sd}) is a consequence of an
important feature of the map (\ref{gsm}), i.e., its \emph{translational
invariance in the }$L$\emph{\ direction} (with period $2\pi $). Other
implications of this invariance are considered below.

\begin{center}
\textbf{B. Quantum Kicked Rotor and Dynamical Localization}
\end{center}

The quantum map for (\ref{gkr}), analogous to the classical one (\ref{gsm}),
is given by the evolution operator in one time period, from $t^{\prime }=t-0$
to $t^{\prime }=t+1-0$:
\begin{equation}
\hat{U}=\exp [-i\hat{L}^{2}/(2\hbar )]\exp [-ikV(\theta )/\hbar ],  \label{U}
\end{equation}
where $\hbar $ is a dimensionless scaled Planck constant and $\hat{L}$ is
the angular-momentum operator $\hat{L}=-i\hbar d/d\theta $ with eigenvalues $
n\hbar $, $n$ integer. It is natural to compare the time evolution of the
expectation value $\left\langle \hat{L}^{2}/2\right\rangle _{t}=\left\langle
\phi ^{(t)}|\hat{L}^{2}/2|\phi ^{(t)}\right\rangle $, $\left\vert \phi
^{(t)}\right\rangle =U^{t}\left\vert \phi ^{(0)}\right\rangle $, with the
classical diffusive evolution expected from (\ref{D}) in the global chaotic
regime (large $k$). First numerical experiments \cite{qkr} indicated that
for sufficiently small $\hbar $ (semiclassical regime) $\left\langle \hat{L}
^{2}/2\right\rangle _{t}$ mimics the classical chaotic diffusion up to some
break-time $t=t_{\mathrm{B}}$ but for $t>t_{\mathrm{B}}$ $\left\langle \hat{L
}^{2}/2\right\rangle _{s}$ stabilizes, becoming bounded in time.

To understand the latter phenomenon, it is necessary to investigate the
nature of the eigenvalues and eigenstates of $\hat{U}$. Since $\hat{U}$ is a
unitary operator, its eigenvalues must lie on the unit circle in the complex
plane and can be written as $\exp (-i\omega )$, where the real quantity $
\omega $ is the so-called \textquotedblleft quasienergy\textquotedblright\
(QE) and ranges in the interval $0\leq \omega <\,2\pi $. \ The QE eigenvalue
problem for $\hat{U}$ is then: 
\begin{equation}
\hat{U}\left\vert \Psi _{\omega }\right\rangle =\exp (-i\omega )\left\vert
\Psi _{\omega }\right\rangle ,  \label{qee}
\end{equation}
where $\left\vert \Psi _{\omega }\right\rangle $ are the QE eigenstates. 
It was shown \cite{sf} that Eq. (\ref{qee}), written in the
angular-momentum representation $\left\langle n|\Psi _{\omega }\right\rangle 
$, is exactly equivalent to the eigenvalue problem of a tight-binding chain
of \textquotedblleft sites\textquotedblright\ $n$, with an on-site potential
which is pseudorandom for generic, irrational values of $\hbar /\left( 2\pi
\right) $. By assuming that such a pseudorandom disorder is effectively
similar to a truly random one which causes Anderson localization, one would
conclude that all the eigenstates $\left\langle n|\Psi _{\omega
}\right\rangle $ are exponentially localized in angular-momentum space and
the QE spectrum $\omega $ is then discrete. These conclusions were
extensively verified numerically. A discrete QE spectrum implies
quasiperiodic and bounded quantum motion, as observed in the first numerical
experiments mentioned above. This bounded quantum motion is known as 
\emph{\textquotedblleft dynamical localization\textquotedblright }. It was
experimentally observed using atom-optics methods, i.e., ultracold atoms
\textquotedblleft kicked\textquotedblright\ by an optical standing wave \cite{e1}.

An important consequence of the exponential localization of QE eigenstates
is that the distribution of the QE level spacings is \emph{Poisson} \cite{ff}. 
This is due to the fact that most of the exponentially localized
eigenstates have far-separated centers on the infinite angular-momentum
space, $-\infty <n\hbar <\infty $, and therefore do not overlap, i.e., they
are uncorrelated.

It was also found later \cite{ds} that in a semiclassical regime of small $\hbar $ the exponential localization length $\xi $\ of the QE eigenstates is roughly
proportional to the diffusion coefficient (\ref{D}). This relation
between $\xi $ and $D$\ is a significant quantum signature of classical
chaos and it was experimentally verified using again atom-optics methods 
\cite{e2}. These experiments also showed that in parameter regimes where
classical accelerator modes exist and chaotic superdiffusion occurs (see
Sec. IIA) the quantum angular-momentum distributions acquire
\textquotedblleft shoulders\textquotedblright , i.e., their initial decay is
slower than the exponential one in the case of normal chaotic diffusion.

\begin{center}
\textbf{C. Quantum Kicked Particle and the Quasimomentum}
\end{center}

In the atom-optics experiments mentioned in Sec. IIB, the quantum KR is
actually realized as a kicked-\emph{particle} system, since atoms move on
lines and not on circles like rotors. However, the quantum kicked particle
can be exactly and simply related to quantum KRs as follows \cite{kp,dd}.
The one-period evolution operator for the quantum kicked particle is given
by Eq. (\ref{U}) with $\hat{L}$ replaced by $\hat{p}$ (linear-momentum
operator) and with $\theta $ replaced by $\hat{x}$ (position operator): 
\begin{equation}
\hat{U}=\exp \left( -i\hat{p}^{2}/(2\hbar )\right) \exp \left[ -ikV(\hat{x}
)/\hbar \right] .  \label{Uo}
\end{equation}
The QE eigenvalue problem for (\ref{Uo}) in the $x$ representation is: 
\begin{equation}
\hat{U}\Psi _{\omega }(x)=\exp (-i\omega )\Psi _{\omega }(x).  \label{qep}
\end{equation}
The $2\pi $-periodicity of (\ref{Uo}) in $\hat{x}$ implies that $\Psi
_{\omega }(x)$ can be chosen to have the Bloch form: 
\begin{equation}
\Psi _{\omega }(x)=\exp (i\beta x)\psi _{\beta ,\omega }(x),  \label{bf}
\end{equation}
where $\beta $ is the \emph{quasimomentum} ($0\leq \beta <1$), whose meaning
is explained below, and $\psi _{\beta ,\omega }(x)$ is $2\pi $-periodic in $
x $. After inserting (\ref{bf}) into Eq. (\ref{qep}), one easily finds that $
\psi _{\beta ,\omega }(x)$ is an eigenstate of 
\begin{equation}
\hat{U}_{\beta }=\exp [-i(\hat{p}+\beta \hbar )^{2}/(2\hbar )]\exp [-ikV(
\hat{x})/\hbar ]  \label{ub}
\end{equation}
with eigenvalue $\exp (-i\omega )$. Due to the $2\pi $-periodicity of $\psi
_{\beta ,\omega }(x)$, one can interpret $x$ as an angle $\theta $ and $\hat{
p}$ in Eq. (\ref{ub}) as an angular-momentum operator $\hat{L}$ with
eigenvalues $n\hbar $. Then, $\hat{U}_{\beta }$ is the evolution operator of
a \textquotedblleft $\beta $-KR\textquotedblright\ and $\beta $ is conserved
during the evolution. To illustrate this conservation and the physical
meaning of $\beta $, assume an initial momentum state $\left\langle
x|p\right\rangle =\exp (ipx/\hbar )$. This can be written in the form 
(\ref{bf}) as $\exp (i\beta x)\exp (inx)$, where $\beta $ and $n$ are,
respectively, the fractional and integer parts of $p/\hbar $ [similarly,
after replacing $\hat{p}$ in Eq. (\ref{ub}) by $\hat{L}$, $\hat{L}+\beta
\hbar $ is the decomposition of the linear-momentum operator into an
\textquotedblleft integer\textquotedblright\ part, $\hat{L}$, and a
conserved \textquotedblleft fractional\textquotedblright\ part, $\beta \hbar 
$]. Then, after $t$ kicks, this state will evolve to a state having still
the form (\ref{bf}) with the same $\beta $: $\left\langle x|p\right\rangle
_{t}=\exp (i\beta x)\phi _{\beta }(x;t)$, where $\phi _{\beta }(x;t)$ is $
2\pi $-periodic in $x$. The usual quantum KR in Sec. IIB corresponds to $
\beta =0$. One can show that the eigenvalue problem for (\ref{ub}) with
general $\beta $ is also equivalent to a tight-binding chain with an on-site
potential that is pseudorandom for generic irrational $\hbar /(2\pi )$.
Dynamical localization is then expected to occur.

Consider now an arbitrary wave packet $\Phi (x)$ of the kicked particle.
This can be always expressed as a superposition of Bloch functions: 
\begin{equation}
\Phi (x)=\int_{0}^{1}d\beta \exp (i\beta x)\phi _{\beta }(x),  \label{phb}
\end{equation}
where $\phi _{\beta }(x)=\sum_{n}\widetilde{\Phi }(n+\beta \hbar )\exp (inx)/
\sqrt{2\pi }$ and $\widetilde{\Phi }(p)$ is the momentum representation of $
\Phi (x)$. Using (\ref{phb}) and the fact that $\hat{p}\exp \left(
i\beta x\right) =\exp \left( i\beta x\right) (\hat{p}+\beta \hbar )$, one then gets
the basic relation 
\begin{equation}
\hat{U}^{t}\Phi (x)=\int_{0}^{1}d\beta \exp (i\beta x)\hat{U}_{\beta
}^{t}\phi _{\beta }(x).  \label{KPR}
\end{equation}
Relation (\ref{KPR}) connects the quantum dynamics of the kicked particle
with that of all the $\beta $-KRs.\newline

\begin{center}
\textbf{D. Quantum Resonance, Antiresonance, Diffusion, and Ratchet
Accelerator}
\end{center}

Classically, the KR map (\ref{gsm}) is translationally invariant in $L$ with
period $2\pi $. Quantally, one may ask under which condition the quantum map
(\ref{ub}) is invariant under a translation in $\hat{p}=\hat{L}$. For a
quantum rotor, such a translation is given by the operator $\hat{T}_{\bar{q}
}=\exp (-i\bar{q}\theta )$, where $\bar{q}$ must be integer since $\theta $
is an angle; thus, $\hat{T}_{\bar{q}}$ is a translation by $\bar{q}\hbar $
in $\hat{L}$ in accordance with the fact that $\hat{L}/\hbar $ has integer
eigenvalues. Using the last fact, the requirement $[\hat{U}_{\beta },\ \hat{T
}_{\bar{q}}]=0$, with $\hat{x}\rightarrow \theta $, $\hat{p}\rightarrow \hat{
L}$ in (\ref{ub}), leads to the conditions \cite{dd}: 
\begin{equation}
\frac{\hbar }{2\pi }=\frac{l}{q},  \label{he}
\end{equation}
\begin{equation}
\beta =\frac{r}{gl}-\frac{gq}{2}\ \mathrm{mod}(1),  \label{br}
\end{equation}
where $l$ and $q$ are coprime integers, $g=\bar{q}/q$ is integer, and $r$
and $gl$ are also coprime integers. For given $(l,q)$, $\beta $ in Eq. 
(\ref{br}) can take any rational value $\beta _{\mathrm{r}}$ in $[0,1)$ since $g$
(or $\bar{q}=gq$) can always be chosen so that $r=(\beta _{\mathrm{r}
}+gq/2)gl$ is integer. Given $\beta =\beta _{\mathrm{r}}$, one chooses $g$
as the smallest positive integer satisfying the latter requirement; in
general, $g>1$. For the usual KR ($\beta _{\mathrm{r}}=0$), $g=1$ if $lq$ is
even and $g=2$ if $lq$ is odd; compare with previous works \cite{cs,fmi}. One
denotes $\beta _{\mathrm{r}}$ by $\beta _{r,g}$, where the integer $r=(\beta
_{\mathrm{r}}+gq/2)gl$ labels all the different values of $\beta _{\mathrm{r}
}$ for given minimal $g$.

The QE states $\psi _{\beta ,\omega }$ for $\beta =\beta _{r,g}$ can be
chosen as simultaneous eigenstates of $\hat{U}_{\beta }$ and $\hat{T}_{gq}$: 
$\hat{U}_{\beta }\psi _{\beta ,\omega }=\exp (-i\omega )\psi _{\beta ,\omega
}$, $\hat{T}_{gq}\psi _{\beta ,\omega }=\exp (-igq\alpha )\psi _{\beta
,\omega }$, where $\alpha $ is a \textquotedblleft
quasiangle\textquotedblright\ varying in the \textquotedblleft Brillouin
zone\textquotedblright\ (BZ) $0\leq \alpha <2\pi /(gq)$. One may view the
Bloch function $\exp (i\beta x)\psi _{\beta ,\omega }(x)$ as a state on the
\textquotedblleft quantum torus\textquotedblright\ $0\leq x<2\pi $, $0\leq
p<gq\hbar $, with toral boundary conditions \cite{id3} specified by $(\alpha
,\beta )$. Using standard methods \cite{cs,fmi}, it is easy to show from the
last two eigenvalue equations that at fixed $\alpha $ one has precisely $gq$
QE eigenvalues $\omega _{b}(\alpha ,\beta )$, $b=0,\dots ,gq-1$. Since $g$
is minimal, the BZ is maximal for the given value of $\beta =\beta _{r,g}$.
Then, as $\alpha $ is varied continuously in the BZ, the $gq$ eigenvalues
generically form $gq$ QE bands with corresponding band eigenstates $\psi
_{b,\alpha ,\beta }$. This band continuous QE spectrum for $\beta =\beta
_{r,g}$ implies an asymptotic ballistic increase in time of the expectation
value of the kinetic energy in any initial wave-packet: $\left\langle \hat{L}
^{2}/2\right\rangle _{\beta ,t}\propto t^{2}$. This is the 
\emph{\textquotedblleft quantum-resonance\textquotedblright } (QR) phenomenon for
rational $\hbar /(2\pi )$ and $\beta $, in contrast to dynamical
localization for irrational $\hbar /(2\pi )$. This QR generalizes the original 
QR for $\beta =0$ \cite{fmi,is}.

An important case is that of the main QRs, $\hbar /(2\pi )=l$ ($q=1$). In
this case, several additional exact results can be derived \cite{dd} since
the evolution operator (\ref{ub}) (with $\hat{x}\rightarrow \theta $, $\hat{p
}\rightarrow \hat{L}$) reduces, up to a nonrelevant constant phase factor,
to 
\begin{equation}
\hat{U}_{\beta }=\exp \left( -i\hbar _{\beta }\hat{L}\right) \exp \left[
-ikV(\theta )/\hbar \right] ,  \label{Ubm}
\end{equation}
where $\hbar _{\beta }=\pi l(2\beta +1)$. We identify $\hat{U}_{\beta }$ in
Eq. (\ref{Ubm}) as the one-period evolution operator for the well-known
linear KR \cite{lkr}, which is exactly solvable for arbitrary potential $
V(\theta )=\sum_{m}V_{m}\exp (-im\theta )$. The following exact results were
obtained \cite{dd}. First, the $gq=g$ QE bands for $\beta =\beta _{r,g}$ are
all nonflat (have finite width) only if there is at least one nonzero
Fourier coefficient $V_{m}$ of $V(\theta )$ with $m$ multiple of $g$
(including $m=\pm g$); then, QR indeed takes place. Otherwise, all the $g$
bands are \emph{flat} (infinitely degenerate) and QR is replaced by a
diametrically opposite phenomenon, \emph{\textquotedblleft quantum
antiresonance\textquotedblright }\ (QAR), a bounded periodic time evolution
of $\left\langle \hat{L}^{2}/2\right\rangle _{\beta ,t}$. For the origin of
the term \textquotedblleft antiresonance\textquotedblright , see Sec. III.
Second, the expectation value of the kinetic energy $\left\langle \hat{p}
^{2}/2\right\rangle _{t}$ of the kicked particle in any initial wave-packet 
(\ref{phb}) grows \emph{linearly (\textquotedblleft
diffusively\textquotedblright )} in time: $\left\langle \hat{p}
^{2}/2\right\rangle _{t}\propto t$, see also other works \cite{kp,dd1}. A similar
growth in time is exhibited by the kinetic energy of an incoherent mixture
of $\beta $-KRs: $\int_{0}^{1}d\beta F(\beta )\left\langle \hat{L}
^{2}/2\right\rangle _{\beta ,t}\propto t$, where $F(\beta )$ is a generic
normalized distribution, $\int_{0}^{1}d\beta F(\beta )=1$. These phenomena
have been experimentally observed for QRs of order $q=1,2,3$ using
atom-optics methods with Bose-Einstein condensates (BECs) \cite{e3}.

In view of the QR ballistic behavior $\left\langle \hat{L}
^{2}/2\right\rangle _{\beta ,t}\propto t^{2}$, one may ask whether, under
some conditions, $\left\langle \hat{L}\right\rangle _{\beta ,t}\propto t$.
The latter behavior may be viewed as a quantum \emph{ratchet acceleration}
since the force $-dV/d\theta $ is unbiased, i.e., it obviously has zero
average, $\int_{0}^{2\pi }d\theta dV/d\theta =0$ (see also Sec. IVD).
Ratchet effects can arise only if some symmetry is broken. The asymmetry is
usually associated with the system, i.e., $V(\theta )$ is asymmetric.
However, a different new kind of asymmetry was proposed \cite{id4}: 
Assume that both $V(\theta )$ and the initial-state amplitude $
\left\vert \phi _{\beta }(\theta )\right\vert $ have inversion symmetry
around \emph{different} symmetry centers; then, the noncoincidence of the
symmetry centers is a \emph{relative asymmetry} which can produce a ratchet
effect. For example, in the simple case of $V(\theta )=\cos (\theta -\gamma
) $ and $\phi _{\beta }(\theta )=[1+\exp (-i\theta )]/\sqrt{4\pi }$, the
symmetry centers of $V(\theta )$ and $\left\vert \phi _{\beta }(\theta
)\right\vert $ are located at $\theta =\gamma $ and $\theta =0$,
respectively.\ One then finds for the main QRs ($q=1$) that $\Delta
\left\langle \hat{L}\right\rangle _{\beta ,t}=\left\langle \hat{L}
\right\rangle _{\beta ,t}-\left\langle \hat{L}\right\rangle _{\beta ,0}$ is
given by \cite{e4}:
\begin{equation}
\Delta \left\langle \hat{L}\right\rangle _{\beta ,t}=\frac{k}{2}\frac{\sin
(\hbar _{\beta }t/2)}{\sin (\hbar _{\beta }/2)}\sin [\hbar _{\beta
}(t+1)/2-\gamma ].  \label{are}
\end{equation}
Now, if $\beta $ takes any of the $l$ resonant values $\beta _{r,1}=r/l-1/2$ 
$\mathrm{mod}(1)$, $r=0,\dots ,l-1$ [from Eq. (\ref{br}) with $g=1$, the
only resonant value of $g$ in the case of $V(\theta )=\cos (\theta -\gamma )$,
see above], one gets from Eq. (\ref{are}) that $\Delta \left\langle \hat{L}
\right\rangle _{\beta ,t}=-k\sin (\gamma )t/2$. The latter QR ratchet effect
was experimentally observed by atom-optics methods for $l=1$ ($\beta
_{r,1}=0.5$) using BECs with quasimomentum width $\Delta \beta \approx 0.1$ 
\cite{e4}. Because of (\ref{are}), this width causes a saturation of the
linear increase of $\Delta \left\langle \hat{L}\right\rangle _{\beta ,t}$.
Theoretical results concerning this saturation effect were also
experimentally confirmed \cite{e4}.

\begin{center}
\textbf{E. Staggered-Ladder QE Spectra and their Quantum-Transport
Manifestations}
\end{center}

We have seen in Sec. IID that under the QR conditions (\ref{he}) and 
(\ref{br}) (with minimal $g$) the QE spectrum of the $\beta $-KR consists of $gq$
bands $\omega _{b}(\alpha ,\beta )$, $b=0,\dots ,gq-1$. It was recently
shown \cite{id5} that this set of bands is actually a 
\emph{\textquotedblleft staggered ladder\textquotedblright }, i.e., the
superposition of $q$ equally-spaced ladders, each consisting of $g$ bands.
Writing $b$ as $b=(c,d)$, where $c$, $c=1,\dots ,q$, labels the $q$ ladders and 
$d$, $d=0,\dots ,g-1$, labels the $g$ bands in each ladder, ladder
$c$ is given by 
\begin{equation}
\omega _{c,d}(\alpha ,\beta )=\omega _{c,0}(\alpha ,\beta )+2\pi dl(\beta
+q/2)\ \text{mod}(2\pi ).  \label{sl}
\end{equation}
This staggered-ladder QE spectrum is another consequence of the
translational invariance of the classical map (\ref{gsm}) in the $L$
direction with period $2\pi $. In the limit of irrational $\beta $ ($
g\rightarrow \infty $), each of he $q$ ladders (\ref{sl}) covers densely the
entire QE range $[0,2\pi )$. We notice that for the usual KR ($\beta =0$),
there are either no ladders ($g=1$ for $lq$ even) or trivial ladders with
spacing $\Delta \omega =\pi $ ($g=2$ for $lq$ odd). Thus, $\beta =0$ is a
nongeneric case for rational $\hbar /(2\pi )$. For the main QRs ($q=1$), the
QE spectrum is just one ladder which, for potentials $V(\theta )$ with a
finite number of harmonics and for sufficiently high-order rational $\beta $,
consists of $g$ flat bands \cite{dd} (see also Sec. IID). The
regularity of the generic staggered-ladder QE spectrum is fundamentally
different from that of the Poisson QE spectrum for irrational $\hbar /(2\pi
) $ (see Sec. IIB).

The spectra (\ref{sl}) have several quantum-transport manifestations \cite{id5}): 
(a) A suppression of QRs for rational $ \beta $ as $g$ increases. 
(b) A dynamical localization for irrational $\beta 
$ which is basically different from that for irrational $\hbar /(2\pi )$;
for example, its time evolution features traveling-wave components in
position space and a staggered-ladder frequency spectrum symmetric around a
central ladder which is independent of the nonintegrability strength $k$.
Most of these phenomena were shown to persist when averaged over realistic
quasimomentum widths $\Delta \beta $ of BECs and should therefore be
experimentally observable.

\begin{center}
\textbf{III. MODULATED KICKED ROTOR AND QUANTUM ANTIRESONANCE}
\end{center}

A significant extension of the KR was introduced \cite{desg},
following previous papers \cite{es,des} in which a special case of this
extension was studied. This is the modulated KR (MKR), defined by a
generalization of the Hamiltonian (\ref{gkr}):
\begin{equation}
H=\frac{L^{2}}{2}+kV(\theta )\sum_{j=0}^{M-1}c_{j}\Delta (t^{\prime }-t_{j}),
\label{mkr}
\end{equation}
where $\Delta (t^{\prime })=\sum_{t=-\infty }^{\infty }\delta (t^{\prime
}-t) $, $c_{j}$ ($j=0,\dots ,M-1$) are $M$ arbitrary coefficients, and $
t_{j} $ ($j=0,\dots ,M-1$) are $M$ \textquotedblleft
times\textquotedblright\ which are arbitrary except of the conditions $0\leq
t_{j}<t_{j+1}\leq 1$ and $t_{0}=0$; one also defines $t_{M}=1$. The
classical map for (\ref{mkr}) can be easily written and generally exhibits
the KAM scenario as in the case of the usual standard map ($M=1$, see Sec.
IIA). In analogy to (\ref{U}), the quantum one-period evolution operator for
the MKR is given by
\begin{equation}
\hat{U}=\prod\limits_{j=0}^{M-1}\exp [-i\tau _{j}\hat{L}^{2}/(2\hbar )]\exp
[-ic_{j}kV(\theta )/\hbar ],  \label{UM}
\end{equation}
where $\tau _{j}=t_{j+1}-t_{j}$ and the factors under the product sign in
(\ref{UM}) are arranged from right to left in order of increasing $j$. Thus, $
\hat{U}$ in Eq. (\ref{UM}) is the composition of $M$ quantum maps $\hat{U}
_{j}$, each corresponding to a KR with time period $\tau _{j}$ and kick
strength $c_{j}k$. Now, the condition for the main QRs of this KR is:
\begin{equation}
\hbar \tau _{j}=4\pi m_{j},\ \   \label{tc}
\end{equation}
where $m_{j}$ is an arbitrary positive integer. Since $\hat{L}/\hbar $ has
integer eigenvalues, condition (\ref{tc}) implies that $\exp [-i\tau _{j}
\hat{L}^{2}/(2\hbar )]=1$ is identically satisfied, so that $\hat{U}
_{j}=\exp [-ikc_{j}V(\theta )/\hbar ]$, which indeed leads to QR for KR $j$.
Then, the MKR is described by the evolution operator 
\begin{equation}
\hat{U}=\prod\limits_{j=0}^{M-1}\exp [-ikc_{j}V(\theta )/\hbar ]=\exp \left[
-i\sum_{j=0}^{M-1}c_{j}kV(\theta )/\hbar \right] ,  \label{UMr}
\end{equation}
leading to QR for the entire MKR system unless 
\begin{equation}
\sum_{j=0}^{M-1}c_{j}=0,  \label{cc}
\end{equation}
implying that 
\begin{equation}
\hat{U}=1  \label{U1}
\end{equation}
identically. Eq. (\ref{U1}) means that all the QE spectrum consists of one
infinitely degenerate level (flat band) $\omega =0$ and \emph{no wave-packet
moves}. This phenomenon is diametrically opposite to that of the QRs
exhibited by the individual KRs \textquotedblleft
composing\textquotedblright\ the MKR according to Eq. (\ref{UMr}). This is
the quantum antiresonance (QAR) phenomenon already considered in Sec. IID.
As far as we are aware, the term \textquotedblleft
antiresonance\textquotedblright\ was first coined in a work \cite{id6} studying
a different class of systems (to be considered in Sec. IV), when referring 
to a yet unpublished paper \cite{des}. This term
reflects the \textquotedblleft cancellation\textquotedblright\ of QRs of
sub-systems composing a given system, due to some condition like (\ref{cc}).
Such a cancellation effect can be shown to be responsible also to QARs
associated with more than one infinitely degenerate QE level, such as the
QAR considered in Sec. IID. One should also mention that the QAR for the $
M=2 $ MKR with $\tau _{0}=\tau _{1}=1/2$, $c_{0}=-c_{1}=1$, and $V(\theta
+\pi )=-V(\theta )$ was shown \cite{desg} to be exactly equivalent to a
well-known period-$2$ QAR \cite{is} occurring in the KR.

It is natural to ask about the behavior of the MKR in the immediate vicinity
of QAR, i.e., when $\hbar \rightarrow \hbar (1+\epsilon )$ in Eq. (\ref{tc})
and condition (\ref{cc}) still holds. It was shown \cite{desg} that this
perturbation of $\hbar $ removes the infinite degeneracy of the single QE
level $\omega =0$, and the QE spectrum is then given by $\omega =\epsilon 
\bar{\omega}$; here $\bar{\omega}$ are the eigenvalues of a one-dimensional
Schr\"{o}dinger equation for the integrable system of a \textquotedblleft
pendulum\textquotedblright\ with a potential $k_{\mathrm{eff}
}^{2}(dV/d\theta )^{2}$, where $k_{\mathrm{eff}}$ is $k$ multiplied by some
quantity dependent on the coefficients $c_{j}$. The corresponding QE
eigenstates are then exponentially localized in angular-momentum space with
a localization length $\xi $, where $\xi ^{-1}$ is not smaller than the
smallest distance of a singularity of $dV/d\theta $ in the complex $\theta $
plane from the real axis.

One can expect that the MKRs should exhibit a rich variety of classical and
quantum transport phenomena associated with many different choices of the
large number $2M$ of parameters $\tau _{j}$ and $c_{j}$, especially when
these parameters are not subjected to conditions like (\ref{tc}) and/or 
(\ref{cc}). In fact, several new phenomena were discovered by Paul Brumer and
co-workers \cite{pb1,pb2,pb3} using just MKRs and modified MKRs whose $\tau
_{j}$ values are all the same ($\tau _{j}=1/M$) and with $c_{j}=\pm 1$.\
These phenomena are: (a) A quantum diffusion, taking place over long time
scales, which is faster than the classical anomalous one (superdiffusion due
to accelerator-mode islands) \cite{pb1}. (b) Controlled enhancement of the
dynamical localization length \cite{pb2}. (c) Classical chaotic ratchet
acceleration, with clear quantum-transport manifestations, exhibited by an
asymmetric MKR with accelerator-mode islands \cite{pb3} (see also Sec. IVD).

The $M=2$ MKR with arbitrary $\tau _{0}=1-\tau _{1}$, $c_{0}$, and $c_{1}$
is the so-called \textquotedblleft double KR\textquotedblright , extensively
studied by several groups during the last decade. A detailed review of the
many results concerning this and related systems is beyond the scope of the
present paper. We refer the interested reader to representative sets of
works \cite{dkr1,dkr2} on this subject.\newpage

\begin{center}
\textbf{IV. NON-KAM SYSTEMS AND\ KICKED HARPER\ MODELS}\\[0pt]
{\ }\\[0pt]
\textbf{A. Classical Kicked Charges in a Magnetic Field and Transport on
Stochastic Webs}
\end{center}

We now consider a class of kicked systems fundamentally different from those
in Secs. II and III. These are charged particles periodically kicked in a
direction perpendicular to a uniform magnetic field $\mathbf{B}$ 
\cite{sw,id7}. We follow here the general approach introduced in our 
work \cite{id7} assuming, for definiteness and without loss of generality, 
that the particles have unit mass and unit charge. Using dimensionless scaled coordinates and notation similar to that in Eq. (\ref{gkr}), the Hamiltonian of the system is:
\begin{equation}
H=\frac{\mathbf{\Pi }^{2}}{2}+kV(x)\sum_{t=-\infty }^{\infty }\delta
(t^{\prime }-t),  \label{kcm}
\end{equation}
where $\mathbf{\Pi }=(\Pi _{x},\Pi _{y})=\mathbf{p}-\mathbf{B}\times \mathbf{
r}/(2c)$ is the kinetic momentum, $\mathbf{p}=(p_{x},p_{y})$ is the
canonical momentum, $\mathbf{B}$ is in the $z$ direction, $\mathbf{r}=(x,y)$,
and $V(x)$ is a general $2\pi $-periodic potential. It is well known 
\cite{jl}\ that the natural degrees of freedom in a uniform magnetic field are
the conjugate pairs $(\Pi _{x},\Pi _{y})$ and $(x_{\mathrm{c}},y_{\mathrm{c}
})$ (coordinates of the center of a cyclotron orbit). Defining $u=\Pi
_{x}/\Omega $ and $v=\Pi _{y}/\Omega $, where $\Omega =B/c$ is the cyclotron
angular velocity, one has the relation $x_{\mathrm{c}}=x+v$, easily
derivable from simple geometry. The Hamiltonian (\ref{kcm}) can thus be
written as
\begin{equation}
H=\frac{\Omega ^{2}}{2}(u^{2}+v^{2})+kV(x_{\mathrm{c}}-v)\sum_{t=-\infty
}^{\infty }\delta (t^{\prime }-t).  \label{kho}
\end{equation}
Since the conjugate mate $y_{\mathrm{c}}$ of $x_{\mathrm{c}}$ is absent in
(\ref{kho}), $x_{\mathrm{c}}$ is a \emph{constant of the motion}. Then, since
also $(u,v)$ are conjugate, (\ref{kho}) is just the Hamiltonian of a
harmonic oscillator periodically kicked by a potential $V(x_{\mathrm{c}}-v)$
dependent on the \textquotedblleft parameter\textquotedblright $x_{\mathrm{c
}}$. The classical map on the $(u,v)$ phase plane for (\ref{kho}), from $
t^{\prime }=t-0$ to $t^{\prime }=t+1-0$, can be easily derived from Hamilton
equations $\dot{u}=\Omega ^{-1}\partial H/\partial v$, $\dot{v}=-\Omega
^{-1}\partial H/\partial u$: 
\begin{equation}
z_{t+1}=[z_{t}+kf(x_{\mathrm{c}}-v_{t})]e^{-i\Omega },  \label{M}
\end{equation}
where $z=u+iv$ and $f(x)=-\Omega ^{-1}dV/dx$. The map (\ref{M}) in the
special case of $x_{\mathrm{c}}=0$ and $V(x)=\cos (x)$ was first presented
by Zaslavsky \textit{et al.} \cite{sw} and it is known as the 
\emph{\textquotedblleft Zaslavsky map\textquotedblright } or 
\emph{\textquotedblleft web map\textquotedblright }. It was later generalized to
the form (\ref{M}) by Dana and Amit \cite{id7}. The main motivation for this
generalization is the sensitivity of the dynamics to the value of $x_{
\mathrm{c}}$. Before discussing this sensitivity, we first consider some
more basic aspects of the system. The harmonic oscillator is a degenerate
system since $\partial ^{2}H_{0}/\partial J^{2}=0$; here $H_{0}=\Omega
^{2}(u^{2}+v^{2})/2=\Omega J$ is the unperturbed Hamiltonian in (\ref{kho})
and $J$ is the action. This in contrast with $H_{0}=L^{2}/2$ in (\ref{gkr}),
with $\partial ^{2}H_{0}/\partial J^{2}=1\neq 0$ ($J=L$). Thus, unlike 
(\ref{gkr}), the KAM theorem cannot be applied to (\ref{kho}), which is
therefore a \emph{non-KAM} system. One then expects that global chaos, i.e.,
unbounded chaotic motion of $(u,v)$, may exist for arbitrarily small $k$ ($
k_{\mathrm{c}}=0$). In fact, such an unbounded motion under the map (\ref{M}) for all $k$ is observed to take place diffusively on a 
\emph{\textquotedblleft stochastic web\textquotedblright } for resonance
(rational) values of $\Omega =2\pi l/n$ ($l$ and $n$ are coprime integers).
For $n=3,4,6$, the web has crystalline symmetry (triangular, square,
hexagonal) while for other values of $n>4$ it has quasicrystalline symmetry.

The chaotic diffusion on the stochastic web for $\Omega =2\pi l/n$ is
characterized by the diffusion coefficient \cite{id7}
\begin{equation}
D(x_{\mathrm{c}})=\lim_{t\rightarrow \infty }\frac{\left\langle \left\vert
z_{nt}-z_{0}\right\vert ^{2}\right\rangle }{2nt},  \label{Dxc}
\end{equation}
where $\left\langle \cdot \right\rangle $ denotes average over an ensemble $
\left\{ (u_{0},v_{0})\right\} $ of initial conditions in the stochastic web.
Analytical and numerical results for crystalline webs indicate a strong
dependence of $D(x_{\mathrm{c}})$ on $x_{\mathrm{c}}$ \cite{id7}. Since a
general ensemble of charged particles exhibits all values of $x_{\mathrm{c}}$, 
a (weighted) average of $D(x_{\mathrm{c}})$ over $x_{\mathrm{c}}$ is
usually necessary. It was shown \cite{id7} that such averaging removes much
of the rich structure (e.g., oscillations) of $D$ versus $k$ at fixed $x_{
\mathrm{c}}$.

For crystalline webs, featuring translational invariance in the $(u,v)$
phase plane, \textquotedblleft accelerator-mode\textquotedblright\ periodic
orbits exist for sufficiently large $k$ \cite{id8}: $u_{mn}=u_{0}+2\pi w_{1}$,
$v_{mn}=v_{0}+2\pi w_{2}$, where $m$ is the minimal period and $
(w_{1},w_{2})$ are integers, not both zero. As in the case of KAM systems
(see Sec. IIA), stable accelerator-mode orbits cause the anomalous chaotic
transport of superdiffusion \cite{id8}: $\left\langle \left\vert
z_{mnt}-z_{0}\right\vert ^{2}\right\rangle \propto t^{\mu (x_{\mathrm{c}})}$, 
$1<$ $\mu (x_{\mathrm{c}})<2$; the anomalous-diffusion exponent $\mu (x_{
\mathrm{c}})$ is again strongly dependent on $x_{\mathrm{c}}$. It has been
suggested \cite{id8} that the strong variation of $D(x_{\mathrm{c}})$ or $
\mu (x_{\mathrm{c}})$ with $x_{\mathrm{c}}$ may be used to \textquotedblleft filter\textquotedblright\ from a general ensemble of charged particles a 
sub-ensemble having any desired well-defined value of $x_{\mathrm{c}}$.

\begin{center}
\textbf{B. Generalized Kicked Harper Models as Realistic Systems}
\end{center}

We now consider a well known quantum-chaos system, the kicked Harper model
(KHM) \cite{sw,khm,ls,gkp,khmr,id9,id10,kh,dfw}, whose most generalized
version is described by the Hamiltonian \cite{id9}:
\begin{equation}
H_{\mathrm{KHM}}=kV_{1}(v)+kV_{2}(u)\sum_{t=-\infty }^{\infty }\delta
(t^{\prime }/2-t),  \label{khm}
\end{equation}
where $V_{1}(v)$ and $V_{2}(u)$ are general (not necessarily periodic)
functions of the phase-space variables $u$ and $v$ defined above. The
original version of the KHM, which appeared in the paper by Zaslavsky 
\textit{et al.} \cite{sw}\textit{,} is the very special case of (\ref{khm})
with $V_{1}(v)=\cos (v)$ and $V_{2}(u)=\cos (u)$. This KHM and its
asymmetric variant with $V_{2}(u)=A\cos (u)$ ($A\neq 1$) were later studied
as kicked-rotor systems \cite{ls,khmr}, i.e., by viewing $u$ as an angle and 
$v$ as an angular momentum. Such systems were found to exhibit a variety of
quantum-transport phenomena for different values of the parameters $k$, $A$,
and a scaled Planck constant. For irrational values of the latter constant,
these phenomena include dynamical localization, quantum diffusion (see also
Sec. IVC), and ballistic quantum motion. The latter two phenomena are not
exhibited by the quantum kicked rotor (see Sec. IIB).

Due to the unusual form of the \textquotedblleft kinetic
energy\textquotedblright\ $kV_{1}(v)$ in (\ref{khm}) (a generally non-quadratic function of the \textquotedblleft momentum\textquotedblright\ $v$), one may ask to what extent the KHM represents a realistic system. The original KHM in the paper
of Zaslavsky \textit{et al.} \cite{sw} was claimed to describe \emph{only
approximately} the system (\ref{kho}) in the case of $\Omega =\pi /2$
(square crystalline web), $x_{\mathrm{c}}=0$, and $V(x)=\cos (x)$. Actually,
it was later shown \cite{id9} that the generalized KHM (\ref{khm}) is 
\emph{exactly related, both classically and quantally}, to the system (\ref{kho})
with $\Omega =\pi /2$, arbitrary $x_{\mathrm{c}}$, and $V(x)$ replaced by a
time-periodic potential $V(x,t^{\prime })$ with period $T=4$ and satisfying
some conditions. To see the classical relation, consider the map $\mathcal{M}
(k)$ for the latter system, from $t^{\prime }=t-0$ to $t^{\prime }=t+4-0$.
This is is the composition of four maps, $\mathcal{M}(k)=\mathcal{M}_{3}
\mathcal{M}_{2}\mathcal{M}_{1}\mathcal{M}_{0}$, where, similarly to (\ref{M}) 
for $\Omega =\pi /2$,
\begin{equation}
\mathcal{M}_{t}\text{:}\ \ v_{t+1}=-[u_{t}+kf(x_{\mathrm{c}}-v_{t},t)],\
u_{t+1}=v_{t},  \label{Mt}
\end{equation}
with $f(x,t^{\prime })=-\Omega ^{-1}\partial V(x,t^{\prime })/\partial x$.
The map $\mathcal{M}_{\mathrm{KHM}}(k)$ for (\ref{khm}), from $t^{\prime
}=2t-0$ to $t^{\prime }=2(t+1)-0$, is:
\begin{equation}
\mathcal{M}_{\mathrm{KHM}}(k)\text{:}\ \ v_{t+1}=v_{t}+kf_{2}(u_{t}),\
u_{t+1}=u_{t}-kf_{1}(v_{t+1}),  \label{mkh}
\end{equation}
where $f_{j}(x)=-\Omega ^{-1}dV_{j}/dx$, $j=1,2$. Then, a straightforward
but tedious calculation \cite{id9} shows that the maps $\mathcal{M}(k)$ and $
\mathcal{M}_{\mathrm{KHM}}(k)$ are exactly related,
\begin{equation}
\mathcal{M}(k)=\mathcal{M}_{\mathrm{KHM}}^{-2}(-k),  \label{mer}
\end{equation}
provided the following conditions are satisfied:
\begin{eqnarray}
V(x_{\mathrm{c}}-v,0) &=&V(x_{\mathrm{c}}+v,2)=V_{1}(v),  \label{c1} \\
V(x_{\mathrm{c}}+u,1) &=&V(x_{\mathrm{c}}-u,3)=V_{2}(u).  \label{c2}
\end{eqnarray}
The quantum version of the exact relation (\ref{mer}) turns out to be 
\cite{id9}: 
\begin{equation}
\hat{U}(k)=-\hat{U}_{\mathrm{KHM}}^{-2}(-k),  \label{qer}
\end{equation}
where
\begin{eqnarray}
\hat{U}(k) &=&-e^{-ikV(x_{\mathrm{c}}-u,3)/\hbar }e^{-ikV(x_{\mathrm{c}
}+v,2)/\hbar }  \notag \\
&&\times e^{-ikV(x_{\mathrm{c}}+u,1)/\hbar }e^{-ikV(x_{\mathrm{c}
}-v,0)/\hbar }  \label{Uk}
\end{eqnarray}
is the evolution operator from $t^{\prime }=t-0$ to $t^{\prime }=t+4-0$ for
the system of kicked charges defined above and $\hat{U}_{\mathrm{KHM}
}(k)=\exp [-kV_{1}(v)/\hbar )]\exp [-kV_{2}(u)/\hbar )]$ is the evolution
operator from $t^{\prime }=t-0$ to $t^{\prime }=t+2-0$ for (\ref{khm}). The
conditions for the validity of (\ref{qer}) are again (\ref{c1}) and (\ref{c2}). 
The minus sign after the equality sign in (\ref{qer}) and (\ref{Uk}) is
of pure quantum origin and is physically irrelevant. If one wishes to
consider only time-independent potentials $V(x,t^{\prime })=V(x)$,
conditions (\ref{c1}) and (\ref{c2}) imply that $V(x)$ must be an \emph{even}
function around $x=x_{\mathrm{c}}$; for example, $V(x)=\cos (x-\gamma )$
with $\gamma =x_{\mathrm{c}}$. The KHM is then symmetric, $V_{1}(x)=V_{2}(x)$. 
It is clear from (\ref{c1}) and (\ref{c2}) that one can \emph{always} find
a time-periodic potential $V(x,t^{\prime })$ which realizes \emph{any} given
generalized KHM. The kicked harmonic oscillator with $V(x)=\cos (x)$ has
already been experimentally realized using atom-optics methods with BECs 
\cite{ht}, but the parameters used do not correspond to $\Omega =\pi /2$,
i.e., to the symmetric KHM.

\begin{center}
\textbf{C. Quantum Antiresonance and Diffusion}
\end{center}

Quantum transport has been studied in general systems (\ref{kho}) 
\cite{id6,dd2,ss}, not related to KHMs, i.e., $\Omega \neq \pi /2$ and/or
conditions (\ref{c1}) and (\ref{c2}) are not satisfied. A first rigorous
result is as follows \cite{id6}. Let us denote by $\hat{U}_{\Omega }$ the
one-period evolution operator for (\ref{kho}) from $t^{\prime }=t-0$ to $
t^{\prime }=t+1-0$ and by $\rho =[\hat{u},\hat{v}]/(2\pi i\Omega )$ a
dimensionless Planck constant. We ask under precisely which conditions the
system will exhibit quantum antiresonance (QAR), i.e., $\hat{U}_{\Omega
}^{m} $ is identically equal to a phase factor for some finite power $m$
[compare with the $m=1$ case of Eq. (\ref{U1}) in Sec. III]. The answer is
that QAR will occur if and only if three conditions are satisfied: (a) The
potential $V(x_{\mathrm{c}}-v)$ is odd, $V(x_{\mathrm{c}}+v)=-V(x_{\mathrm{c}
}-v)$, up to some additive constant. (b) There is classical
\textquotedblleft crystalline\textquotedblright\ resonance, i.e., $\Omega
=2\pi l/n$ (see Sec. IVA) with either $n=4$ (square crystalline case) or $
n=6 $ (hexagonal crystalline case); in both cases, the power $m=n$. (c) $
\rho $ is integer for $n=4$ while $\sqrt{3}\rho /2$ is integer for $n=6$. It
was also shown \cite{id6} that for $n=4$ the QAR is due to
the\textquotedblleft cancellation\textquotedblright\ of the main QRs of two
KHMs. This is analogous to the QAR of MKRs \cite{desg}, arising from the
cancellation of the main QRs of KRs (see Sec. III).

In a second work \cite{dd2}, basic aspects of the QE spectrum and quantum
transport were studied as functions of $x_{\mathrm{c}}$ and $\rho $ for $n=4$
and $V(x)=-\cos (x)$. It was shown that if the parameter $\varepsilon =k\sin
(\pi \rho )/(2\pi \rho )$ is sufficiently small, $\varepsilon \ll 1$ and $
\left\vert \varepsilon /[2\cos (x_{\mathrm{c}})]\right\vert \ll 1$ (i.e., $
x_{\mathrm{c}}$ is not very close to $\pi /2$), the QE spectrum at fixed $x_{
\mathrm{c}}$ and $k/\rho $ is approximately the spectrum of a (symmetric)
Harper Hamiltonian \cite{hm} $\hat{H}=-2\cos (x_{\mathrm{c}})[\cos (\hat{u}
)+\cos (\hat{v})]$. The latter spectrum as function of $\rho $, $0\leq \rho
\leq 1$, forms the famous \textquotedblleft Hofstadter
butterfly\textquotedblright\ (HB) \cite{hb}. However, for $x_{\mathrm{c}
}=\pi /2$, one has $\hat{H}=0$: the potential $V(x_{\mathrm{c}}-v)=-\sin (v)$ is
odd and QAR occurs for $\rho =1$ (see above), with an infinitely degenerate
QE spectrum. In fact, the QE spectrum for $x_{\mathrm{c}}=\pi /2$ at fixed $
k/\rho $ shrinks to a point as $\rho \rightarrow 0$ or $\rho \rightarrow 1$.
After scaling the QEs by the $\rho $-dependent factor $\varepsilon ^{-1}$,
the spectral structure as function of $\rho$ at fixed small value of $k/\rho $ becomes very close to that of a \textquotedblleft double HB\textquotedblright .

It is known \cite{gkp1}\ that the spectrum of the Harper Hamiltonian is
fractal (a Cantor set) for generic, irrational $\rho $ and that such
spectrum leads to a \emph{\textquotedblleft quantum
diffusion\textquotedblright } of the expectation value $\left\langle \hat{u}
^{2}\right\rangle _{t}$ or $\left\langle \hat{v}^{2}\right\rangle _{t}$ for
asymptotically large times $t$. This should be then also the approximate
behavior of the expectation value of the kinetic energy in (\ref{kho}), at
least for sufficiently small values of the parameter $\varepsilon $ above:
\begin{equation}
\left\langle \frac{\Omega ^{2}}{2}(\hat{u}^{2}+\hat{v}^{2})\right\rangle
_{t}\approx D_{\mathrm{q}}(x_{\mathrm{c}})t,  \label{Dqxc}
\end{equation}
where $D_{\mathrm{q}}(x_{\mathrm{c}})$ is the quantum-diffusion coefficient.
The asymptotic behavior (\ref{Dqxc}) was extensively verified numerically
and the approximate formula $D_{\mathrm{q}}(x_{\mathrm{c}})\approx D_{
\mathrm{q}}(0)\cos (x_{\mathrm{c}})$, valid at least in simple cases, was
derived \cite{dd2}. As expected, $D_{\mathrm{q}}(x_{\mathrm{c}})\rightarrow
0 $ in the QAR limit of $x_{\mathrm{c}}\rightarrow \pi /2$. The
quantum-transport behavior (\ref{Dqxc}) for the non-KAM system (\ref{kho})
is in sharp contrast with the dynamical localization for KAM (KR) systems
(see Secs. IIB and III). The quantum diffusion of a general ensemble of
charged particles, exhibiting all values of $x_{\mathrm{c}}$, is
characterized by a (weighted) average of $D_{\mathrm{q}}(x_{\mathrm{c}})$
over $x_{\mathrm{c}}$ \cite{dd2}, in analogy to the classical case (see Sec.
IVA).

It was already noted in the work mentioned above \cite{id6} (see Fig. 1 there) 
that while QAR occurs in a strong quantum regime of non-small $\rho \gtrsim 1$, it has a distinct classical analogue: For small $k$, the chaotic diffusion on the
crystalline stochastic web for odd potential $V(x_{\mathrm{c}}-v)$ [say,
$V(x)=-\cos (x)$ and $x_{\mathrm{c}}=\pi /2$] is \emph{much slower} than that
for even potential $V(x_{\mathrm{c}}-v)$ [say, $V(x)=-\cos (x)$ and $x_{
\mathrm{c}}=0$]. This classical phenomenon was explained in a later work in
the case of $n=4$ \cite{prk}: For even potential, relation (\ref{mer})
holds, so that the map $\mathcal{M}(k)$ is, like the map $\mathcal{M}_{
\mathrm{KHM}}(k)$ in (\ref{mkh}), a perturbation of order $O(k)$ of the
identity map $v_{t+4}=v_{t}$, $u_{t+4}=u_{t}$; on the other hand, for odd
potential, the map $\mathcal{M}(k)$ turns out to be a much smaller
perturbation, of order $O(k^{2})$, of the identity map.

\begin{center}
\textbf{D. Quantum-Chaotic Ratchet Accelerators}
\end{center}

Classical Hamiltonian ratchets are systems in which a directed current can
emerge in the chaotic region from an unbiased force, i.e., a force whose
phase-space and/or time average is zero. This current is a mean position
velocity (usual velocity) or momentum velocity (acceleration). In the latter
case, one has a ratchet accelerator, first introduced by Gong and Brumer 
\cite{pb3} using MKRs in a strong-chaos regime (see end of Sec. III; see
also the very recent ratchet accelerator for arbitrarily weak chaos 
\cite{id11}). For a persistent Hamiltonian ratchet effect to occur, it is
necessary that the system is asymmetric and possesses transporting stability
islands (accelerator-mode islands in the case of ratchet accelerators) whose
total flux is non-zero due to the asymmetry. Thus, such an effect cannot
arise in a fully chaotic regime (with no stability islands) even if the
system is asymmetric.

On the other hand, \emph{quantum-ratchet-acceleration} effects can 
\emph{generically} occur also in systems whose classical phase space is 
\emph{fully chaotic} \cite{qra,id12,id13}. This was first shown by Gong and Brumer 
\cite{qra} using an asymmetric KHM (\ref{khm}) which is fully chaotic and
does not exhibit a classical ratchet effect. For generic irrational values
of a scaled Planck constant in a semiclassical regime, a strong quantum
ratchet acceleration is observed using a spatially uniform initial state (a
zero-momentum state). This acceleration is due to the existence of cantori
(broken KAM\ tori) in the \textquotedblleft vertical\textquotedblright\
(momentum $v$) direction whose broken phase-space structures (gaps in the
KAM tori) have typical size smaller than the scaled Planck constant. Thus,
quantally, the gaps appear as \textquotedblleft closed\textquotedblright\
and the cantori act effectively as vertical KAM tori which cause, already
classically, a ratchet acceleration. This strong quantum effect was found to
be \emph{robust to noise} \cite{qra} and should therefore be experimentally
realizable using kicked harmonic oscillators that are exactly equivalent to
KHMs (see Sec. IVB).

In two recent papers \cite{id12,id13}, the quantum ratchet acceleration in
fully chaotic KHMs was studied for small rational values of the
dimensionless Planck constant $\rho $ (using the notation in Sec. IVC), $
\rho =1/N$ ($N\gg 1$). This corresponds to high-order QRs in a deep
semiclassical regime. The most important difference between this study and
the one above of Gong and Brumer is that the initial states used are 
\emph{maximally uniform in phase space} (MUPS), not only in ordinary space. 
Each such state corresponds to a \emph{\textquotedblleft
von-Neumann\textquotedblright  lattice} in phase space whose unit cell ($0\leq
u<a$, $0\leq v<b$) has Planck area $ab=h=4\pi ^{2}\rho =4\pi ^{2}/N$; the lattice origin is labeled by phase-space \emph{quasicoordinates} $(w_{1},w_{2})$ ranging 
in the Planck cell, $0\leq w_{1}<a$, $0\leq w_{2}<b$. It was shown that a
MUPS state $\left\vert w_{1},w_{2}\right\rangle $ leads to a QR ratchet
acceleration $I(w_{1},w_{2})$ much stronger than that obtained by using pure
momentum states, already in \emph{completely symmetric} KHMs. As in the case
of the symmetric QR ratchets described in Sec. IID, this effect is due to
the \emph{relative asymmetry} associated with the noncoincidence of the symmetry
centers of the KHM and the MUPS state $\left\vert w_{1},w_{2}\right\rangle $. 
It was also shown that the distribution of $I(w_{1},w_{2})$ over $
\left\vert w_{1},w_{2}\right\rangle $ is a Gaussian with mean $\left\langle
I\right\rangle =0$ and variance $\left\langle I^{2}\right\rangle =2D/N^{2}$,
where $D$ is the classical chaotic-diffusion coefficient in the momentum ($v$) direction. Besides $\left\vert w_{1},w_{2}\right\rangle $, other initial
states were considered which approximate $\left\vert
w_{1},w_{2}\right\rangle $ to some arbitrary order denoted by $B$ ($0\leq
B\leq \infty $); here $B=0$ and $B=\infty $ correspond, respectively, to
pure momentum states and MUPS states $\left\vert w_{1},w_{2}\right\rangle $.
It was found that the quantum ratchet acceleration over the approximating
states has zero mean, $\left\langle I\right\rangle _{B}=0$, and variance $
\left\langle I^{2}\right\rangle _{B}$ increasing monotonically with $B$: 
$\left\langle I^{2}\right\rangle _{0}$ for pure momentum states is
significantly smaller than $\left\langle I^{2}\right\rangle _{\infty
}=2D/N^{2}$ for the MUPS states. For sufficiently low order $B$, the
approximating states should be experimentally realizable.

\begin{center}
\textbf{V. CONCLUSIONS}
\end{center}

In this brief review, we have focused on representative classes of 1D kicked
systems with a periodic time dependence associated entirely with the kicking
potential. These systems have been extensively investigated during the last
three decades. Less is known about kicked systems in more than one dimension
and/or with a non-periodic time dependence. For the benefit of the
interested reader, we conclude by considering very briefly two important
examples of the latter systems. The first example is a rotor kicked by a
potential quasiperiodic in time with three incommensurate frequencies
(including the frequency of the periodic delta function) \cite{cgs}. For
irrational values of a scaled Planck constant, the QE problem for this
system is equivalent to that of a 3D pseudorandom tight-binding model
(compare with the 1D case in Sec. IIB). Then, as a nonintegrability
parameter is increased, there occurs a transition from pseudorandom Anderson
localization to unbounded diffusion (delocalized regime) at a critical value
of the parameter. This transition and related phenomena have been
experimentally observed using atom-optics methods \cite{e5}.

The second example is that of a 1D periodically kicked particle subjected to
an additional constant force or linear potential 
\cite{id4,qam,qam1,qam2,qam3,qam4,qam5,qam6}. The theoretical interest in this
system began following the experimental discovery of the \textquotedblleft
quantum accelerator modes\textquotedblright\ (QAMs) in the free-falling
frame (FFF) of periodically kicked atoms falling under gravity \cite{qam}.
In the FFF Hamiltonian, the linear potential does not appear but the
momentum, in the kinetic-energy term, increases linearly in time 
\cite{id4,qam3}; this makes the relevant time dependence of the system
non-periodic. The QAMs in the FFF were theoretically explained \cite{qam3}
as associated with a vicinity of the main QRs, i.e., with values $\hbar
=2\pi l+\epsilon $ ($l$ integer, see Sec. IID) of the scaled Planck constant 
$\hbar $. It was shown that this vicinity defines a \textquotedblleft
quasiclassical\textquotedblright\ regime in which $\epsilon $ plays the role
of a fictitious Planck constant and the quantum evolution in the FFF can be
approximately described by a classical map. Then, a wave packet initially
trapped in an accelerator-mode island of this map \textquotedblleft
accelerates\textquotedblright ; this is a QAM. The experimentally observed
robustness of QAMs under relatively large deviations $\epsilon $ from $\hbar
=2\pi l$ was explained, in the framework of the quasiclassical
approximation, as a \textquotedblleft mode-locking\textquotedblright\
phenomenon \cite{qam4,qam5,qam6}. Theoretical predictions were verified by
several experiments \cite{qam2,qam4}.

\end{document}